\documentclass[a4paper, 12pt]{article}

\usepackage{changepage}

\usepackage[utf8]{inputenc}
\usepackage[T1]{fontenc}

\usepackage[]{graphicx}
\usepackage{caption}
\usepackage{subcaption}

\usepackage{epsf,amsmath,bbold,amsfonts,stmaryrd}
\usepackage{mathrsfs}
\usepackage{tikz}
\usetikzlibrary{arrows,matrix,positioning}
\usepackage{appendix}
\usepackage{amssymb}
\usepackage{float}
\usepackage{color} 
\usepackage{dsfont}
\usepackage{bm}
\usepackage[colorlinks]{hyperref}
\hypersetup{pageanchor=false,citecolor=red,urlcolor=red}
\usepackage{indentfirst}
\usepackage{slashed}
\usepackage[numbers,square,comma,compress,merge]{natbib}
\usepackage{soul}

\usepackage[skip=2pt,font=small]{caption}

\allowdisplaybreaks

\def\a{\alpha}
\def\b{\beta}

\def\d{\delta}

\def\f{\frac}
\def\g{\gamma}

\def\G{\Gamma}

\def\l{\left}

\def\m{\mu}
\def\n{\nu}

\def\p{\partial}

\def\r{\right}
\def\s{\sigma}

\def\x{\xi}

\def\z{\zeta}

\def\be{\begin{equation}}
\def\ee{\end{equation}}

\def\bes{\begin{equation*}}
\def\ees{\end{equation*}}

\def\bea{\begin{eqnarray}}
\def\eea{\end{eqnarray}}

\def\ba{\begin{array}}
\def\ea{\end{array}}

\def\bc{\begin{center}}
\def\ec{\end{center}}

\def\bl{\begin{flushleft}}
\def\el{\end{flushleft}}

\def\br{\begin{flushright}}
\def\er{\end{flushright}}

\def\bi{\begin{itemize}}
\def\ei{\end{itemize}}

\def\bt{\begin{tabular}}
\def\et{\end{tabular}}

\newtheorem{question}{Question}
\def\bq{\begin{question}}
\def\eq{\end{question}}

\newtheorem{definition}{Def}
\def\bd{\begin{definition}}
\def\ed{\end{definition}}

\newtheorem{answer}{Answer}
\def\ban{\begin{answer}}
\def\ean{\end{answer}}

\newtheorem{possibleanswer}{Possible answer}
\def\bpa{\begin{possibleanswer}\normalfont}
\def\epa{\end{possibleanswer}}

\newtheorem{theorem}{Theorem}
\def\bth{\begin{theorem}}
\def\eth{\end{theorem}}

\usepackage{esvect}

\usepackage{slashed}

\usepackage{xspace}
\newcommand*{\ie}{i.e., }
\newcommand*{\eg}{e.g., }
\newcommand*{\rhs}{rhs\@\xspace}

\newcommand*{\Eq}{Eq.\@\xspace}
\newcommand*{\Eqs}{Eqs.\@\xspace}

\usepackage{xifthen}
\newcommand*\diff{\mathrm{d}} 
\newcommand*\ldiff[2][]{ \ifthenelse{\isempty{#1}}{ \diff #2}{\diff^#1#2} \,} 
\let\limitint\int 
\renewcommand{\int}{\limitint \!} 

\newsavebox\myboxA
\newsavebox\myboxB
\newlength\mylenA

\newcommand*\xoverline[2][0.75]{%
    \sbox{\myboxA}{$\m@th#2$}%
    \setbox\myboxB\null
    \ht\myboxB=\ht\myboxA%
    \dp\myboxB=\dp\myboxA%
    \wd\myboxB=#1\wd\myboxA
    \sbox\myboxB{$\m@th\overline{\copy\myboxB}$}
    \setlength\mylenA{\the\wd\myboxA}
    \addtolength\mylenA{-\the\wd\myboxB}%
    \ifdim\wd\myboxB<\wd\myboxA%
       \rlap{\hskip 0.5\mylenA\usebox\myboxB}{\usebox\myboxA}%
    \else
        \hskip -0.5\mylenA\rlap{\usebox\myboxA}{\hskip 0.5\mylenA\usebox\myboxB}%
    \fi}
\makeatother

\begin{document}

\begin{titlepage}

\vspace*{-2.5cm}
\begin{adjustwidth}{}{-.45cm}
\br
{
\begin{tabular}{@{}l@{}}
\small LMU--ASC~18/21\\
 \small FTPI--MINN--21--10 \\
 \small UMN--TH--4017/21
 \end{tabular}
 }
\er
\end{adjustwidth}

\vspace*{.5cm}

\begin{center}
\bf \Large{Matter matters in Einstein-Cartan gravity}
\end{center}

\begin{adjustwidth}{-1.3cm}{-.7cm}
\begin{center}
\textsc{Georgios K. Karananas,$^{\star}$~Mikhail Shaposhnikov,$^\dagger$\\
Andrey Shkerin,$^{\ddagger}$~Sebastian Zell\,$^\dagger$}
\end{center}
\end{adjustwidth}

\begin{center}
\it {$^\star$Arnold Sommerfeld Center\\
Ludwig-Maximilians-Universit\"at M\"unchen\\
Theresienstra{\ss}e 37, 80333 M\"unchen, Germany\\
\vspace{.4cm}
$^\dagger$Institute of Physics\\
Laboratory of Particle Physics and Cosmology\\
\'Ecole Polytechnique F\'ed\'erale de Lausanne (EPFL)\\ 
CH-1015, Lausanne, Switzerland\\
\vspace{.4cm}
$\ddagger$William I. Fine Theoretical Physics Institute\\
School of Physics and Astronomy\\
University of Minnesota \\
Minneapolis, MN 55455, USA
}
\end{center}

\begin{center}
\small
\texttt{georgios.karananas@physik.uni-muenchen.de}\\
\texttt{mikhail.shaposhnikov@epfl.ch}\\
\texttt{ashkerin@umn.edu}\\
\texttt{sebastian.zell@epfl.ch}
\end{center}

\vspace{1cm}

\begin{abstract}
We study scalar, fermionic and gauge fields coupled nonminimally to gravity in the Einstein-Cartan formulation. We construct a wide class of models with nondynamical torsion whose gravitational spectra comprise only the massless graviton. Eliminating non-propagating degrees of freedom, we 
derive an equivalent theory in the metric formulation of gravity. It features contact interactions of a certain form between and among the matter and gauge currents. We also discuss briefly the inclusion of curvature-squared terms.
\end{abstract}

\end{titlepage}

\section{Introduction}

There is overwhelming evidence in favor of General Relativity (GR) as the theory of classical gravity. Nevertheless, this leaves open a far-reaching question about the choice of fundamental fields. Different options lead to different formulations of gravity. One possibility, which is most commonly used, is the metric approach. In this formulation, the metric is selected as the only fundamental field whereas the Christoffel symbols are defined \emph{a priori} as functions of the metric and are fixed to correspond to the Levi-Civita connection. This implies that the gravitational dynamics is fully captured by curvature. 

Another formulation of GR is provided by the Einstein--Cartan (EC) theory~\cite{Cartan:1922,*Cartan:1923,*Cartan:1924,*Cartan1925, Einstein1928,*Einstein19282}. In this case, the vielbein and  spin connection assume the role of fundamental fields, from which the metric and  Christoffel symbols can be subsequently derived. Since the connection is independent of the metric, the theory features torsion in addition to curvature. 
While the metric and EC formulations of gravity look very different, they are exactly equivalent in the pure GR case. The way to see this is as follows. In EC gravity, it is possible to solve for the connection. If no matter is included, the result is the Levi-Civita connection. Hence, torsion still vanishes, but this time dynamically as a consequence of its equations of motion. 

The different formulations represent an inherent theoretical ambiguity contained within GR. A theory of pure gravity cannot distinguish between them, and this puts on equal footing the various choices of fundamental fields, including the most commonly used metric approach. It is important to stress the difference between the formulations of GR and its modifications, such as massive (for a review see~\cite{deRham:2014zqa}) or DGP~\cite{Dvali:2000hr} gravity. The latter can already be distinguished from GR in pure gravity.

The list of formulations of GR which are equivalent in pure gravity is rather long. Let us indicatively mention the ones based on the  Palatini~\cite{Palatini1919, Einstein1925},\footnote{See~\cite{Unzicker:2005in} for an English translation of~\cite{Einstein1928,Einstein19282} and~\cite{Einstein1925}.} affine~\cite{dyson_1925,Einstein_affine,10.2307/20488450,schrodinger1944affine,kijowski1978}, or teleparallel~\cite{Hayashi:1967se,Hayashi:1979qx,Kopczynski_1982,muller1985tetrad} (see~\cite{DeAndrade:2000sf} for a review) gravity. Given this zoo of options, the question arises if one can select a preferred one. The answer is twofold. First, some choices may lead to conceptual advantages. For example, EC gravity follows from gauging the Poincar\'e group~\cite{Utiyama:1956sy, Kibble:1961ba}, which brings gravity closer to the rest of the interactions in Nature; see~\cite{Hehl:1976kj,Blagojevic:2003cg,Obukhov:2006gea,Blagojevic:2012bc,Blagojevic:2013xpa,Obukhov:2018bmf} for reviews. Furthermore, the first-order formalism (where the metric and connection are independent) allows for boundary terms that are well-defined without any need for an infinite counterterm~\cite{Ashtekar:2008jw}. Of course, such arguments are not irrefutable. Second, and more importantly, once matter fields are involved, the ``degeneracy'' may very well be lifted, so that different frameworks lead to different predictions. This can open a way, at least in principle, to distinguish between them via observations and experiment. In order for this to be possible, however, it is necessary to systematically quantify their differences. The goal of the present paper is to contribute to this program. 

In what follows we focus on EC gravity, which encompasses the metric and Palatini\,\footnote{The Palatini formulation treats the metric and affine connection as independent fields, yet the connection is assumed to be symmetric.} versions as special cases. First, we shall discuss EC theory in more detail. In general, gauging of the Poincar\'e
group leads to the introduction of 40 degrees of freedom (in 4 spacetime
dimensions). They are distributed among the 24-component spin connection and
16-component tetrad/vierbein. 
Owing to the local nature of the Poincar\'e transformations,
not all 40 degrees of freedom can be physical. In fact, using the
gauge freedom, half of them can be eliminated. This leaves at most 20 propagating degrees of freedom. It is an easy exercise to decompose the latter into states with
definite spin and parity; this reveals that 12 degrees of freedom are
in the spin-2 sector (the massless graviton + 2 massive tensors), 6 in the spin-1
sector (2 massive vectors), and 2 in the spin-0 sector (2 massive
scalars)~\cite{Neville:1978bk,Sezgin:1979zf}. In EC gravity, however, only the massless graviton is endowed with a kinetic term. This implies that the rest of the gravitational states are not propagating and that the connection is nondynamical. 

Once coupled to matter, the inequivalence of EC and metric gravity manifests itself in two ways. First, since there is no \textit{a priori} assumption about the symmetry of the Christoffel symbols, matter can source torsion even when it is only coupled minimally to gravity. This is \eg the case for fermions. Generically, the resulting effects are suppressed by powers of the Planck mass $M_P$ \cite{Kibble:1961ba, osti_4843429}. Secondly, one can add additional terms to the action. An example consists in the Holst term~\cite{Hojman:1980kv, Nelson:1980ph, Castellani:1991et, Holst:1995pc}, which is the full contraction of the curvature tensor with the totally antisymmetric symbol. In metric gravity this vanishes identically---it actually comes down to the algebraic Bianchi identity of the Riemann tensor---but it gives a nontrivial contribution once torsion is present. The additional terms in the action come with \textit{a priori} undetermined dimensionless coupling constants. If they are bigger than $1$, they lead to effects that are already visible well below the Planck scale.

A consequence of the equivalence between the metric and EC formulations of GR in the absence of matter is that their particle spectrum is identical and comprises the two polarizations of the massless graviton. Interestingly, this continues to be the case also when the theory is coupled to matter sectors. The connection now picks up extra pieces involving the matter fields, but it remains nondynamical. Correspondingly, it is possible to solve for torsion. By plugging the result back into the action, one can derive an equivalent torsion-free theory, in which the matter sector is supplemented with a set of specific higher-dimensional operators. In other words, the EC framework acts as a set of selection rules in that it singles out a particular subset of all possible higher-dimensional operators consistent with the gauge redundancies of the system. 

Over the years there has been a lot of progress in constructing the most general EC theory with (nonminimally) coupled scalar and fermionic fields~\cite{gr-qc/0505081, hep-th/0507253,Alexandrov:2008iy, 0807.2652, 0811.1998, 0902.0957, 0902.2764, Diakonov:2011fs, 1212.0585, Langvik:2020nrs}. So far, the most complete model, which encompasses the previously mentioned works as special cases, was investigated in \cite{Shaposhnikov:2020frq}. However, even in the study \cite{Shaposhnikov:2020frq} numerous terms were (implicitly) excluded from the Lagrangian without justification. Our goal here is to generalize the previous investigations by first proposing systematic criteria for construction an action of matter coupled to EC gravity and then including \emph{all} terms that fulfill these criteria. In doing so,  we take into account fermions, a real scalar, as well as an Abelian Higgs model. Of particular interest is the latter case, for it is a stripped to its bare essentials version of the Standard Model (SM) that nonetheless captures all the salient features of its symbiosis with EC gravity. Among the various terms that appear in the action, there are also couplings of the $U(1)$ scalar current to torsion. For the fully-fledged SM this corresponds to the hypercharge. In the effective metric description this translates into novel higher-dimensional terms describing contact interactions of this current with itself and with the other fields.

The paper is organized as follows. In Sec.~\ref{sec:constr_action}, we introduce some basic concepts of EC theory and then lay out criteria for methodically constructing an action of matter fields coupled to gravity. Based on these principles, we discuss in details all terms that we include in the action. In Sec.~\ref{sec:equiv_met_theory}, we derive the effective metric description of the theory. As a sanity check, we compare various limiting cases of our findings with existing results in the literature. In Sec.~\ref{sec:curvatureSquare}, we discuss how our considerations are altered in the presence of curvature-squared terms. In particular, we introduce and study a model in which the inclusion of such a term does not lead to new propagating degrees of freedom. In Sec.~\ref{sec:conclusion}, we conclude.

\paragraph{Conventions.} Throughout this paper we work in 4 spacetime dimensions. Greek letters are reserved for spacetime indices and capital Latin letters for Lorentz indices. Both the spacetime $g_{\m\n}$ and Minkowski $\eta_{AB}$ metrics have mostly plus signature. Our convention for the gamma matrices is 
\be
\l\{\g_A,\g_B\r\} = - 2 \eta_{AB} \ ,~~~\g_5 =-i\g^0\g^1\g^2\g^3= i \g_0\g_1\g_2\g_3 \ ,
\ee
meaning that 
\be
\l\{\g^A,\Sigma^{BC}\r\}= - 2 \epsilon^{ABCD}\g^5\g_D \ ,~~~\l[\g^A,\Sigma^{BC}\r]= - 4i \eta^{A[B}\g^{C]} \ ,~~~
\Sigma_{AB} \equiv \f i 2 \l[\g_A,\g_B\r] \ .
\ee
The totally antisymmetric tensor is taken such that $\epsilon_{0123}=1$. We work in natural units $c=\hbar=1$.

\pagebreak

\section{Constructing the action}   
\label{sec:constr_action}

\subsection{Geometrical preliminaries}
\label{sec:geom_prel}

The mathematical toolbox of EC gravity is that of the  Poincar\'e gauge theory. The gravitational interaction emerges from gauging the Poincar\'e group~\cite{Utiyama:1956sy, Kibble:1961ba}. In order to make this possible, one needs to introduce more degrees of freedom as compared to GR. Specifically, it is necessary to introduce two gauge fields in order to localize Lorentz transformations and translations. These are the (spin) connection $\omega_\m^{AB}$, which is antisymmetric in its upper indices, and the tetrad/vierbein $e_\m^A$, respectively. They live in the (co)tangent space of the spacetime manifold. The space is endowed with two bases. The one is induced by the spacetime metric $g_{\mu\nu}$; to refer to it, we use Greek indices, covariant under diffeomorphisms. The other is an orthonormal noncoordinate basis referred to with Latin indices and enjoying covariance under local Lorentz transformations. These two bases are connected via the tetrad \eg $V^A = e_\m^A V^\mu$, for a vector $V^\mu$. In particular, the two metrics are related via
\be
g_{\a\b}=e^A_\a e^B_\b \eta_{AB} \ ,~~~\eta_{AB} = e_A^\a e_B^\b g_{\a\b}\ .
\ee
The covariant derivatives in the two bases read 
\be
\label{eq:cov_derivatives}
D_\m V^\a = \partial_\m V^\a + \Gamma^\a_{\sigma\m} V^\sigma \ ,~~~D_\m V^A = \partial_\m V^A + \omega_\m^{AB} V_B \ ,
\ee
with $\G^\kappa_{\m\n}$ the affine connection. They transform homogeneously under diffeomorphism and local Lorentz transformations, respectively. The fact that the coefficients $\Gamma^\a_{\sigma\m}$ and $\omega_\m^{AB}$ correspond to the same connection expressed in different bases leads to $D_\mu e_\n ^ A=0$.\footnote{We remark that it is also possible to consider the case in which $\Gamma^\a_{\sigma\m}$ and $\omega_\m^{AB}$ represent two different connections, and correspondingly $D_\mu e_\n ^ A$ does not vanish \cite{Raatikainen:2019qey}.} This condition ensures the compatibility of the two expressions in \Eq~\eqref{eq:cov_derivatives} and moreover implies
\be 
\label{eq:ChirstoffelFromSpinConnection}
\Gamma^\kappa_{\n\m} = e^\kappa_A\l( \p_\m e^A_\n+\omega^A_{\m B} e^B_\n \r)   \ .
\ee
Thus, the $\Gamma^\kappa_{\n\m}$ can be defined as functions of the tetrad and the spin connection. The antisymmetry of the spin connection, $\omega _ \m ^ {AB} = - \omega _ \m ^ {BA}$, implies metric compatibility
\be
\nabla_\m g_{\a\b} = 0 \qquad \Leftrightarrow \qquad \nabla_\m \eta_{AB} = 0 \ .
\ee

Finally, the field strengths corresponding to the spin connection and tetrad can be obtained by acting with the commutator of covariant derivatives on a vector. This yields the explicit form of the curvature $F_{\m\n}^{AB}$ and torsion $T_{\m\n}^A$:
\bea
\label{eq:curv_def}
& & F_{\m\n}^{AB} = \p_\m \omega_\n^{AB} -\p_\n \omega_\m^{AB}+\omega^A_{\m C}\omega^{CB}_\n - \omega^A_{\n C}\omega^{CB}_\m \ , \\
\label{eq:tors_def}
& & T_{\m\n}^A = \p_\m e_\n^A -\p_\n e_\m^A+ \omega^A _{\m B}e^B_\n -\omega^A _{\n B}e^B_\m  \ .
\eea
Using appropriate Poincar\'e- and diffeomorphism-invariant
combinations of $F$ and $T$, one can write down an effective theory by
expanding in powers of the field strengths, or, equivalently, in the
derivatives of the fields. We will discuss this in details in the following.

\subsection{Selection rules for the terms in the action}
\label{sec:criteria}

In the following, we shall construct an action for matter coupled to gravity in the EC formulation. Thereby our goal is to solely focus on those effects that arise from the ambiguities in the coupling of matter to gravity. Correspondingly, we shall demand equivalence to the metric theory in pure gravity, and also leave the matter sector on its own invariant. This leads us to impose the following three criteria:

\begin{itemize}
   \item[\emph{i)}] The purely gravitational part of the action should solely contain operators of mass dimension not greater than 2.
   \item[\emph{ii)}] In the flat spacetime limit, \ie for $e_\m ^A  =  \d_\m^A,~\omega_\m^{AB} = 0$, the matter Lagrangian should be renormalizable.
   \item[\emph{iii)}] The coupling of matter to gravity should only happen through operators of mass dimension not greater than 4. 
\end{itemize}

Let us elaborate on the significance of the above. The purpose of criterion~\emph{i)} is to ensure that the gravitational sector is equivalent to GR in the metric formulation. A necessary condition to achieve this is to have the same particle content. Hence, we demand that out of the plethora of possible states of gravitational origin, the massless graviton is the only one that propagates. Why this requirement motivates us to only consider terms of mass dimension not greater than 2 can be understood by counting the derivatives~\cite{Diakonov:2011fs}. Torsion contains a derivative of the tetrad, see~\Eq~(\ref{eq:tors_def}), \ie each occurrence of $T$ in the action counts as one derivative. Curvature $F$ consists of derivatives of torsion and torsion-squared terms (see \Eqs~(\ref{eq:curv_def}) and also~(\ref{eq:f_decomp}) below), hence it counts as two derivatives. Now torsion and curvature have mass dimensions 1 and 2, respectively. Consequently, the number of derivatives is equivalent to the mass dimension, and criterion~\emph{i)} is tantamount to restricting ourselves to terms with at most 2 derivatives. As is well-known, operators with more derivatives would generically lead to the appearance of new propagating degrees of freedom. Moreover, some of them would also have kinetic terms with wrong signs.

We have to mention, however, that there are exceptions to this dictum. First, certain combinations of curvature-squared terms result in healthy particle spectra~\cite{Stelle:1977ry,Neville:1978bk,Hayashi:1979wj,*Hayashi:1980qp,Sezgin:1979zf,Sezgin:1981xs,Kuhfuss:1986rb,Yo:1999ex,Yo:2001sy,Nair:2008yh,Nikiforova:2009qr,Karananas:2014pxa,*Karananas:2016ltn,Obukhov:2017pxa,Blagojevic:2017ssv,Blagojevic:2018dpz,Lin:2018awc,Jimenez:2019qjc,Lin:2019ugq,Percacci:2019hxn}. Second, one can devise particular higher-curvature theories which do not propagate any particles apart from the massless graviton. Thus, condition~\emph{i)} is sufficient but not necessary for the absence of new propagating degrees of freedom. We further discuss this point in Sec.~\ref{sec:curvatureSquare}.  

Next, criterion~\emph{ii)} implies that before coupling to gravity, the matter sector only contains terms of mass dimension not bigger than 4. This postulate is crucial for the predictiveness of our setup. As we will show, the model built according to~\emph{i)} and~\emph{ii)} can be equivalently expressed as a torsion-free theory that contains a specific set of higher-dimensional operators of the matter fields. If we were to drop condition~\emph{ii)}, \ie added from the beginning all possible higher-dimensional operators to the action, the inclusion of torsion would not bring any new information. Equivalently, one can say that we use torsion as a criterion to select specific higher-dimensional operators of the matter theory. Needless to say, the soundness of such an approach remains to be checked. One way to do so is to explore its consequences. Insofar it leads to predictions that are consistent and in agreement with observations, this can be regarded as an~\emph{a posteriori} justification for imposing criteria~\emph{i)} and~\emph{ii)}. In the present paper, we shall lay the groundwork for exploring the consequences of the theory defined by these conditions.

Finally, criterion~\emph{iii)} states that also after coupling matter to gravity, the theory only contains terms of mass dimension not bigger than 4. Imposing such a requirement appears to be natural in view of the analogous condition~\emph{ii)} in the matter sector. Like condition~\emph{i)}, it ensures that the operators coupling gravity and matter do not introduce any additional propagating degrees of freedom. However, unlike~\emph{i)} and \emph{ii)}, criterion~\emph{iii)} can be easily relaxed in explicit computations without spoiling the  predictiveness of our setup and without the danger of invoking extra propagating particles. For this reason, our analysis below is, in fact, more general, and our results remain valid even if condition~\emph{iii)} is relaxed.

To summarize, we have proposed certain criteria to construct a generic class of models for coupling matter to gravity in the EC formulation. Postulates~\emph{i)} and~\emph{ii)} are restrictive enough to ensure that the pure matter sector does not contain any higher-dimensional operators and that the pure gravity sector is equivalent to GR in its metric formulation. Thus, any new effects that we discover originate solely from the interaction of matter with gravity. In other words, we explore the consequences of the fact that there is not a unique way of coupling matter to gravity.

\subsection{Decomposition of torsion and contorsion}
\label{sec:decomp}

Due to the antisymmetry of torsion (defined in \Eq~\eqref{eq:tors_def}) in the spacetime indices, it has 24 independent components in 4 dimensions. These can be conveniently grouped into three irreducible pieces: a vector $v_\mu$, a pseudovector $a_\mu$ and the 16-component reduced torsion tensor $\tau_{\mu\n\rho}$. Explicitly,  
\be \label{eq:tors_all}
v_\m = T^\n_{~\m\n} \ ,~~~a_\m = \epsilon_{\m\n\rho\sigma}T^{\n\rho\sigma} \ ,~~~\tau_{\m\n\rho} =\frac 2 3 \l( T_{\m\n\rho} -v_{[\n} g_{\rho]\m} - T_{[\n\rho]\m} \r) \ ,
\ee
with $T_{\m\n\rho}=e_{\mu A} T^A_{\n\rho}$, and summation over repeated indices is tacitly assumed. As customary, square (round) brackets stand for antisymmetrization (symmetrization) of the corresponding indices. The reduced torsion tensor is subject to the following conditions
\be 
\label{eq:tau_constr}
\tau^{\n}_{~\m\n} = 0  \ ,~~~\epsilon_{\m\n\rho\sigma}\tau^{\n\rho\sigma}= 0 \ .
\ee
In terms of its irreducible components, the torsion tensor reads
\be
\label{eq:tors_irreps}
T_{\m\n\rho} = \frac 2 3 v_{[\n}g_{\rho]\m} - \frac 1 6 a^\sigma \epsilon_{\m\n\rho\sigma} +\tau_{\m\n\rho} \ .
\ee

Moreover, we introduce the torsionless spin connection $\mathring{\omega}^{AB}_\m$ which is a function of the tetrad. To find its expression, we demand that the rhs of \Eq~(\ref{eq:tors_def}) vanish. The resulting algebraic equation can be solved for the connection and yields
\be
\label{eq:conn_tors_free}
\mathring{\omega}^{AB}_\m= \frac{1}{2}\l[e ^ {\n A} \l ( \p _ \m e _ \n ^ B - \p _ \n e _ \m ^ B \r )  - e ^ {\n B} \l ( \p _ \m e _ \n ^ A - \p _ \n e _ \m ^ A \r ) 
- e _ {\m C} e ^ {\n A} e ^ {\lambda B} \l ( \p _ \n e _ \lambda ^ C - \p _ \lambda e _ \n ^ C \r ) \r] \ .
\ee
It follows from \Eq \eqref{eq:ChirstoffelFromSpinConnection} that this is equivalent to
\be
\mathring{\Gamma}^\kappa_{\m\n}=\f 1 2 g^{\kappa\lambda}\l( \p_\m g_{\lambda\n} +\p_\n g _{\m\lambda}-\p_\lambda g_{\m\n}\r) \ ,
\ee
where $\mathring{\Gamma}^\kappa_{\m\n}$ are the Christoffel symbols of the torsion-free Levi-Civita connection. 

The full spin connection can be split as
\be
\label{eq:conn_split} 
\omega_{\m}^{AB}=\mathring{\omega}_{\m}^{AB}+C_{\m}^{AB} \ , 
\ee
implying the following decomposition of the affine connection 
\be
\Gamma^\kappa_{\m\n} =  \mathring{\Gamma}^\kappa_{\m\n} + e^\kappa_A e^B_\n C_{\m B}^{A} \ ,
\ee
where we introduced the contorsion tensor $C_\m ^{AB}$. The latter is related to torsion as
\be
\label{eq:contorsion_1}
C_\m ^{AB} = \f 1 2 e^{\a A} e^{\b B} \l(T_{\a\b\m}-T_{\b\a\m}-\textbf{}T_{\m\a\b} \r)  \ .
\ee
Plugging into the above the decomposition~(\ref{eq:tors_irreps}) of torsion in terms of the vector, pseudovector and reduced tensor, we can express contorsion as 
\be
\label{eq:contorsion_2}
C_\m ^{AB} = e^{\a A} e^{\b B}\left( \frac 2 3 v_{[\b}g_{\a]\m} +\frac {1}{12} \epsilon_{\a\b\m\n}a^\n+2\tau_{[\a\b]\m}\right) \ . 
\ee
It is evident from \Eq~\eqref{eq:tors_def} that 
\be 
\label{eq:TorsionFromContorsion}
T_{\m\n}^A = C^A _{\m B}e^B_\n -C^A _{\n B}e^B_\m   \ .
\ee
From Eqs.~(\ref{eq:contorsion_1}) and~(\ref{eq:TorsionFromContorsion}) it follows that torsion and contorsion contain the same information about the spacetime geometry and are completely equivalent from a dynamical point of view.

Having introduced the necessary ingredients and notation, in the following we shall systematically construct the most general action of gravity coupled to matter that fulfills the  conditions~\emph{i)} and~\emph{ii)} spelled out in Sec.~\ref{sec:criteria}. It will become apparent that working in terms of the torsion components $v_\m, a_\m, \tau_{\m\n\rho}$ greatly facilitates the analysis.

\subsection{Pure gravity}
\label{sec:grav}

Let us first discuss a purely gravitational theory. We already mentioned that the restriction to at most two derivatives of the fields implies that the action can only contain terms quadratic in torsion and linear in curvature. Regarding torsion, this leads to the following seven terms~\cite{Diakonov:2011fs}
\be 
\label{eq:torsionSquare}
 \frac{1}{\sqrt{g}}\p_\m\l(\sqrt{g}v^\m\r) \ ,~~~ \frac{1}{\sqrt{g}}\p_\m\l(\sqrt{g}a^\m\r)\ ,~~~v_\mu v^\mu\ ,~~~a_\mu a^\mu \ ,~~~v_\mu a^\mu\ ,~~~\tau_{\m\n\rho}\tau^{\m\n\rho}\ ,~~~\epsilon^{\m\n\rho\sigma}\tau_{\m\n\lambda}\tau_{\rho\sigma}^{~~\lambda} \ ,
\ee
where we denoted $g=-\det(g_{\m\n})$. As for curvature, only two invariants are admissible. These are the parity preserving Einstein-Hilbert  and parity violating Holst terms given by
\be
\label{eq:f_def}
F \equiv \frac {1}{8\sqrt g} \epsilon_{ABCD}\epsilon^{\m\n\rho\sigma}F^{AB}_{\m\n}e^C_\rho e^D_\sigma \ ,~~~\text{and}~~~\tilde F \equiv \f {1}{\sqrt g}\epsilon^{\m\n\rho\sigma}e_{\rho C}e_{\sigma D}F_{\m\n}^{CD}  \ ,
\ee
respectively. Using~\Eqs (\ref{eq:conn_split})-(\ref{eq:contorsion_2}), we can decompose the above into torsion-free and torsionful contributions. This gives
\begin{align}
\label{eq:f_decomp}
F & = \frac{\mathring R}{2} +\frac{1}{\sqrt{g}}\p_\m\l(\sqrt{g}v^\m\r) -\frac 1 3 v_\m v^\m + \frac{1}{48} a_\m a^\m +\frac 1 4 \tau_{\m\n\rho}\tau^{\m\n\rho}\ ,  \\
\label{eq:tilde_f_decomp}
\tilde F & = -\frac{1}{\sqrt{g}}\p_\m\l(\sqrt{g}
a^\m\r)+\frac 2 3 a_\m v^\m -\frac 1 2 \epsilon^{\m\n\rho\sigma}\tau_{\lambda\m\n}\tau^{\lambda}_{~\rho\sigma}\ , 
\end{align}
where the torsion-free Riemannian curvatures are defined as
\be
\mathring R = \mathring R^\m_\m \ ,~~~\mathring R_{\m\n}=\mathring R_{\n\m}= \d^\lambda_\sigma \mathring R^\lambda_{~\m\sigma\n} \ ,~~~\mathring R^\lambda_{~\m\sigma\n} = \p_\sigma \mathring \Gamma^\lambda_{\n\m} -\p_\n \mathring \Gamma^\lambda_{\sigma\m} +\mathring \Gamma^\lambda_{\sigma\rho} \mathring \Gamma^\rho_{\n\m}-\mathring \Gamma^\lambda_{\n\rho}\mathring \Gamma^\rho_{\sigma\m} \ .
\ee
In expanding $\tilde F$, we dropped the term $\propto\epsilon^{\m\n\rho\sigma}\mathring{R}_{\m\n\rho\sigma}$, since it vanishes identically by virtue of the symmetries of $\mathring{R}_{\m\n\rho\sigma}$. Note also that the decompositions~(\ref{eq:f_decomp}) \&~(\ref{eq:tilde_f_decomp}) contain all seven torsion invariants from Eq.~(\ref{eq:torsionSquare}), albeit with fixed coefficients. 

Overall, the action of pure gravity reads
\be
\begin{aligned}
\label{eq:E-H_action_1}
S_{\rm gr} =  M_P^2\int \diff^4 x\sqrt g\Bigg [ F +\frac{1}{4\bar\g} \tilde F &+\frac{\tilde c_{vv}}{2} v_\mu v^\mu  + \tilde c_{va}v_\mu a^\mu + \frac{\tilde c_{aa}}{2}	a_\mu a^\mu\\
&+\tilde c_{\tau\tau}\tau_{\a\b\g} \tau^{\a\b\g} + {\tilde c'}_{\tau\tau} \epsilon^{\m \n \rho \sigma} \tau_{\lambda\m\n} \tau^\lambda_{~\rho\sigma} + 2\Lambda \Bigg ] \ ,
\end{aligned}
\ee
where the $\bar\g$ and $\tilde c$'s are arbitrary dimensionless constants; $\bar\g$ is called the Barbero-Immirzi parameter~\cite{Immirzi:1996dr, Immirzi:1996di}. For completeness, we also included a cosmological constant term  $\Lambda$, although its presence does not play any role in the subsequent analysis. 

To get a better handle on the dynamics of the theory~(\ref{eq:E-H_action_1}), it is useful to express it in its equivalent  metric-only form by integrating out the nondynamical connection $\omega_\m^{AB}$. Although straightforward, this approach quickly becomes algebraically tedious, especially in the presence of matter. Therefore, we will simplify the computation using the following procedure. First, we split the connection as in~(\ref{eq:conn_split}) in a torsion-less part $\mathring \omega_\m^{AB}$ and contorsion $C_\m^{AB}$. Secondly, we use \Eq \eqref{eq:contorsion_1} to replace contorsion by torsion $T_{\a\b\m}$. Thirdly, we split torsion in its irreducible components $v_\mu$, $a_\mu$, $\tau_{\m\n\rho}$ (see \Eq \eqref{eq:tors_irreps}). All these operations are bijective, \ie $\omega_\m^{AB}$ uniquely determines the triplet $v_\mu$, $a_\mu$, $\tau_{\m\n\rho}$ and vice versa. Therefore, varying the action with respect to $\omega_\m^{AB}$ is equivalent to varying with respect to $v_\mu$, $a_\mu$, $\tau_{\m\n\rho}$. Opting for the second option, we will derive the equations of motions for the irreducible components of torsion, solve them and plug the result back into the action. 

Practically, what we just described means that we use~(\ref{eq:f_decomp}) and~(\ref{eq:tilde_f_decomp}) and rewrite the action~(\ref{eq:E-H_action_1}) as
\be
\begin{aligned}
\label{eq:E-H_action_2}
S_{\rm gr} =  M_P^2\int  \diff^4 x\sqrt{g}\Bigg[\frac{\mathring R}{2}   &+\frac{c_{vv}}{2} v_\mu v^\mu  + c_{va}v_\mu a^\mu + \frac{c_{aa}}{2}	a_\mu a^\mu\\
&+c_{\tau\tau}\tau_{\a\b\g} \tau^{\a\b\g} + {c'}_{\tau\tau} \epsilon^{\m \n \rho \sigma} \tau_{\lambda\m\n} \tau^\lambda_{~\rho\sigma}+2\Lambda\Bigg] \ ,
\end{aligned}
\ee
where we dropped the total derivatives of $v_\mu$ and $a_\mu$, and introduced the shifted constants 
\bea
&\displaystyle c_{vv} = \tilde c _{vv} - \frac{2}{3} \ ,~~~c_{va} = \tilde c_{va} +\frac{1}{6\bar\g}\ ,~~~c_{aa} = \tilde c_{aa} + \frac{1}{24} \ ,&\nonumber \\
&\displaystyle c_{\tau\tau} = \tilde c_{\tau\tau} + \frac{1}{4} \ ,~~~{c'}_{\tau\tau} = {\tilde c'}_{\tau\tau} - \frac{1}{8\bar\g} \ .&\label{constants}
\eea
Note that the contorsion contribution is completely factored out and contained in the torsion-square terms. 

Varying~(\ref{eq:E-H_action_2}) w.r.t. $v_\mu,a_\mu,\tau_{\m\n\rho}$, we readily see that torsion (and contorsion) is not sourced and therefore all three quantities vanish. This means that in vacuum the theory is indistinguishable from GR in the metric formulation:
\be
S_{\rm gr} = \int  \diff^4 x\sqrt{g}\Bigg [\frac{M_P^2}{2}\mathring R  +2\Lambda \Bigg] \ .
\ee
This will change once matter is introduced.

Note finally that from Eq.~(\ref{constants}) it follows that the actions~(\ref{eq:E-H_action_1}) and~(\ref{eq:E-H_action_2}) are completely equivalent. However, the second form is preferable from the point of view of computational convenience as well as the fact that all of the constants are independent.

\subsection{Fermions}

We are now in a position to generalize our considerations by coupling matter fields to EC gravity. Let us start with fermions. For the sake of illustration, we focus on a single massless four-component spinor $\Psi$, with the generalization to more generations being straightforward. The action comprising the kinetic term for $\Psi$ and its possible interactions with torsion reads~\cite{Diakonov:2011fs}
\be
\begin{aligned}
\label{eq:ferm_act_1}
&\displaystyle S_{f} = \int \diff^4 x\sqrt{g} \Bigg[ \frac{i}{2} \l(\overline{\Psi}\g^\m \mathring{D}_\m\Psi - \overline{\mathring{D}_\m \Psi}\g^\m \Psi \r)
\\ & \qquad\qquad\qquad\qquad+
\l(\z^v_V V_\m +\z^v_A A_\m\r)v^\m + \l(\z^a_V V_\m +\z^a_A A_\m\r)a^\m \Bigg] .
\end{aligned}
\ee
Here $\g^\m=e^\m_A\g^A$, and  the torsion-free fermionic covariant derivative reads 
\be 
\mathring{D}_\m = \p_\m +\frac 1 8 \mathring{\omega}_\m^{AB}[\g_A,\g_B] \ .
\ee
Further, 
$\zeta^{v}_V$, $\zeta^{a}_V$, $\zeta^{v}_A$ and $\zeta^{a}_A$ are arbitrary coefficients. Finally, 
\be 
V_\m = \bar{\Psi} \gamma_\mu \Psi \;, ~~~ A_\m = \bar{\Psi} \gamma_5 \gamma_\mu \Psi
\ee
are the vector and axial fermionic currents, respectively. 

A few comments are in order here. First, we could have started from a non-canonical kinetic term for the fermion
\be \label{eq:ferm_delta}
S_f \supset \int \diff ^ 4 x \frac{i}{2} \l(\overline{\Psi}\l(1+\d\g_5\r)\g^\m \mathring{D}_\m\Psi - \overline{\mathring{D}_\m \Psi}\l(1+\d\g^5\r)\g^\m \Psi \r) \ ,
\ee
with $\d$ a real constant. But now we can canonically normalize the field, \ie perform a field redefinition such that the kinetic term of the fermion again assumes the form as displayed in the first line of \Eq~\eqref{eq:ferm_act_1}. In fermionic interaction terms, this transformation can be reabsorbed by a rescaling of the coupling constants.  This stays true also if interactions of fermionic currents with $v_\m$ and $a_\m$ are included, provided that all possible contributions are taken into account in the action. Thus, we can omit the term \eqref{eq:ferm_delta} without loss of generality.

Second, one may wonder why we have not included a coupling between $\Psi$ and the reduced torsion tensor, viz $\bar \Psi \g^\m\g^\n\g^\rho\Psi \tau_{\m\n\rho}$. Using the properties of the $\g$-matrices, it is not difficult to show that
\be
\bar \Psi \g^\m\g^\n\g^\rho\Psi \tau_{\m\n\rho} \propto V_\m \tau^\n_{~\m\n} +i A_\m \epsilon^{\m\n\rho\s}\tau_{\n\rho\s} \ .
\ee
Both terms in the \rhs of this expression vanish identically by virtue of the constraints~(\ref{eq:tau_constr}).

Finally, it is worth mentioning that like in the pure gravity case, we could have equally well started with nonminimally coupled fermions~\cite{hep-th/0507253,Alexandrov:2008iy, 1212.0585}
\be
\begin{aligned}
S_{f} &= \int \diff^4 x\sqrt{g} \Bigg[ \frac i 2 \overline{\Psi}(1-i\a-i\b \g^5)\g^\m D_\m\Psi - \frac i 2 \overline{D_\m \Psi}(1+i\a+i\b \g^5)\g^\m \Psi \\
&\quad\qquad\qquad\qquad\qquad\qquad\qquad+ \l(z^v_V V_\m +z^v_A A_\m\r)v^\m + \l(z^a_V V_\m +z^a_A A_\m\r)a^\m \Bigg] \ ,
\end{aligned}
\ee
where the covariant derivative $D_\m$ now includes the full connection. The real constants $\a,\b$ are nonminimal  couplings, and $z^{v/a}_{V/A}$ are analogous to the couplings in Eq.~(\ref{eq:ferm_act_1}). After decomposing the connection as in Eq.~(\ref{eq:conn_split}), one finds that $\a$ and $\b$ feed into the torsion-current interactions, and one ends up with Eq.~(\ref{eq:ferm_act_1}) upon identifying
\be
\z^v_V = z^v_V -\f \a 2 \ ,~~~\z^v_A = z^v_A -\f \b 2 \ ,~~~\z^a_V = z^a_V\ ,~~~\z^a_A = z^a_A -\f 1 8 \ .
\ee
Thus, the nonminimal couplings are not independent parameters once torsion is coupled to the fermionic currents.

\subsection{Real scalar field}
\label{sec:real}

Let us move to the scalar-gravity sector of the EC theory. We first consider a real scalar $\phi$; the case of a complex field is discussed in Sec.~\ref{sec:complex}. We find that the most general gravi-scalar action, at most quadratic in the derivatives of all fields, reads
\be
\begin{aligned}
\label{eq:action_real_scalar}
S_{\rm{gr}+\phi} &= \int \diff^4 x \sqrt{g}\Bigg[\frac{M_P^2}{2}\Omega^2\mathring{R} - \frac {(\p_\m \phi)^2}{2} - U + v^\mu \partial_\mu Z^v + a^\mu \partial_\mu Z^a \\
&+ \f{M_P^2}{2}\Big(G_{vv} v_\mu v^\mu  + 2G_{va}v_\mu a^\mu + G_{aa}	a_\mu a^\mu 
 +G_{\tau\tau}\tau_{\a\b\g} \tau^{\a\b\g}+ \tilde{G}_{\tau\tau} \epsilon^{\m \n \rho \sigma} \tau_{\lambda\m\n} \tau^\lambda_{~\rho\sigma}\Big) \Bigg]\ .
\end{aligned}
\ee
Here $\Omega^2$, $U$, $Z^{v/a}$ and $G_{ij}$ are, in general, arbitrary functions of $\phi$ (``coefficient functions''). The function $\Omega^2$ represents the nonminimal coupling of the  field to the Ricci scalar, and $U$ the potential. We can reduce the (infinite) freedom contained in these functions to a limited number of parameters by imposing condition~\emph{iii)} from Sec.~\ref{sec:criteria}. The latter only permits nonminimal interaction terms which are at most quadratic in the field. Requiring invariance under $\phi\to -\phi$, we find
\be\label{eq:coefffunreal}
\Omega^2=1+\frac{\xi\phi^2}{M_P^2}\;, ~~~ Z^{v/a}=\z^{v/a}_\phi\phi^2 \;, ~~~ G_{ij}= c_{ij} \l(1+\frac{\x_{ij}\phi^2}{M_P^2}\r) \ ,
\ee
where no summation over the repeated $i,j$ indices is implied, $\xi$ is the standard nonminimal coupling constant and $\z^{v/a}_\phi$, $c_{ij}$ and $\x_{ij}$ are also constants. The analysis of this section is carried out for the general coefficient functions; their form (\ref{eq:coefffunreal}) will be used in Sec.~\ref{sec:lim} to compare with the previously studied models. Note also that, although allowed in principle, the terms $\tau^{\m\n}_{~~\,\m}\p_\n Z^\tau(\phi)$ and $\epsilon^{\kappa\lambda\m\n}\tau_{\kappa\lambda\m}\p_\n \tilde{Z}^\tau(\phi)$ coupling the derivative of $\phi$ to the reduced torsion tensor are identically zero due to~(\ref{eq:tau_constr}). 

For completeness, let us mention that, as before, it is possible to start with the field $\phi$ nonminimally coupled  to the curvatures $F$ and $\tilde F$ instead of $\mathring R$:
\begin{equation}
\begin{aligned}
S_{\rm{gr}+\phi} &= \int \diff^4 x \sqrt{g} \Bigg [ M_P^2\Omega^2 F +M_P^2\tilde{\Omega}^2 \tilde F - \frac {(\p_\m \phi)^2}{2} -U + v^\mu \partial_\mu z^v + a^\mu \partial_\mu z^a  \\
&\qquad+ \f{M_P^2}{2}\Big(g_{vv} v_\mu v^\mu  + 2g_{va}v_\mu a^\mu + g_{aa}	a_\mu a^\mu +g_{\tau\tau}\tau_{\a\b\g} \tau^{\a\b\g} + \tilde{g}_{\tau\tau} \epsilon^{\m \n \rho \sigma} \tau_{\lambda\m\n} \tau^\lambda_{~\rho\sigma}\Big) \Bigg]\ ,
\end{aligned}
\end{equation}
with $\tilde{\Omega}^2$, $z^{v/a}$ and $g_{ij}$ arbitrary coefficient functions. Using Eqs.~(\ref{eq:f_decomp}) and~(\ref{eq:tilde_f_decomp}), we end up with the action~(\ref{eq:action_real_scalar}), upon identifying
\bea
&\quad\quad z^v =Z^v+M_P^2\Omega^2 \ ,~~~z^a =Z^a-M_P^2\tilde{\Omega}^2\ ,~~~g_{vv}=G_{vv}+\dfrac{2\Omega^2}{3} \ ,
\\
&g_{va}=G_{va}-\dfrac{2\tilde{\Omega}^2}{3}  \ ,~~g_{aa}=G_{aa}-\dfrac{ \Omega^2}{24}  \ ,~~g_{\tau\tau}=G_{\tau\tau}-\dfrac{\Omega^2}{2}  \ ,~~\tilde g_{\tau\tau}=\tilde G_{\tau\tau}+\tilde{\Omega}^2  \ . 
\eea

\subsection{Complex scalar field and gauge bosons}
\label{sec:complex}

The results of the previous section can be readily generalized to the case of a complex scalar field $\Phi$. For simplicity, we focus on a local $U(1)$ theory and denote by $F_\m$ the corresponding Abelian gauge field. This is enough to capture the differences from the case of  the real scalar. 

The action of the theory reads as follows
\be
\begin{aligned}
\label{eq:action_complex_scalar}
S_{\rm{gr}+\Phi} &= \int \diff^4 x \sqrt{g}\Bigg[\frac{M_P^2}{2}\Omega^2\mathring{R} - (\mathcal D_\m \Phi)^\ast \mathcal D^\m \Phi -U-\frac 1 4 F_{\m\n}^2 \\
&\quad\quad\quad\quad\quad\quad+ v^\mu \partial_\mu Z_\Phi^v + a^\mu \partial_\mu Z_\Phi^a + Z^v_S S_\mu v^\mu + Z^a_S S_\mu a^\mu  \\
&+ \f{M_P^2}{2}\Big(G_{vv} v_\mu v^\mu+ 2G_{va}v_\mu a^\mu  + G_{aa}	a_\mu a^\mu+G_{\tau\tau}\tau_{\a\b\g} \tau^{\a\b\g} + \tilde{G}_{\tau\tau}\epsilon^{\m \n \rho \sigma} \tau_{\lambda\m\n} \tau^\lambda_{~\rho\sigma}\Big) \Bigg]\ .
\end{aligned}
\ee
Here $\Omega^2$, $U$, $Z^{v/a}_{\Phi}$, $Z^{v/a}_S$ and $G_{ij}$ are arbitrary coefficient functions depending on $\Phi^*\Phi$. Next, $\mathcal D_\m = \p_\m -i e F_\m$ corresponds to the $U(1)$-covariant derivative with $e$ the gauge coupling, and $S_\m$ is the scalar Noether current associated with the global part of the $U(1)$ symmetry 
\be
S_\m = -\f{i}{2} \l(\Phi^\ast (\mathcal D_\m \Phi) - (\mathcal D_\m \Phi)^\ast \Phi\r) \ . 
\ee
Further, the field strength is given by
\be 
F_{\m\n}=\p_\m F_\n-\p_\n F_\m \;,
\ee
where it is important to note that partial derivatives are used. In a torsion-free theory, one could have equivalently employed covariant derivatives since the contributions with Christoffel symbols would cancel out. Once torsion is present, however, this is no longer true. In this case, using covariant derivatives in $F_{\m\n}$ would break the $U(1)$ gauge invariance~\cite{Hehl:1976kj}. 

Employing condition~\emph{iii)} from Sec.~\ref{sec:criteria}, we obtain constraints on the coefficient functions analogous to Eq.~(\ref{eq:coefffunreal}),  
\be\label{eq:coefffuncomplex}
\Omega^2=1+\frac{2\xi\Phi^*\Phi}{M_P^2}\;,~~ Z^{v/a}_\Phi=2\Phi^*\Phi\zeta^{v/a}_\Phi \;, ~~ Z^{v/a}_S=\zeta^{v/a}_S\;, ~~ G_{ij}= c_{ij} \l(1+\frac{2\x_{ij}\Phi^\ast\Phi}{M_P^2}\r) \ , 
\ee
where $\xi$, $\zeta^{v/a}_\Phi$, $\zeta^{v/a}_S$, $c_{ij}$ and $\xi_{ij}$ are constants and there is no summation in the $i,j$ indices.

Comparing with the case of the real scalar considered in Sec.~\ref{sec:real}, there are two additional terms in the action, namely the couplings between the $U(1)$ scalar current and the torsion vectors. As we show below, once torsion is eliminated, these give rise to dimension-six contact interactions between $\Phi$, $S_\m$, $V_\m$ and $A_\m$. 

Before moving on, let us point out that non-Abelian groups can be treated in a completely analogous manner. However, it is not possible to form gauge-invariant objects using non-Abelian currents (at least to this order in derivatives), hence the latter cannot be coupled to torsion.

\section{Equivalent metric theory}
\label{sec:equiv_met_theory}

\subsection{Full action}

In the previous section we constructed the most general actions for EC  gravity without and with matter fields, which at the same time satisfy the requirements listed in Sec.~\ref{sec:criteria}. 
The full action of the theory we will consider in what follows reads
\be
\label{eq:action_rho_f_1}
S = S_{\rm{gr}+\Phi}+S_f \ ,
\ee
where $S_{\rm{gr}+\Phi}$ and $S_f$ are given in \Eqs~(\ref{eq:action_complex_scalar}) and~(\ref{eq:ferm_act_1}), respectively. To keep the discussion as general as possible, we choose the complex scalar field action $S_{\rm{gr}+\Phi}$, since it allows for additional interaction terms involving the current $S_\m$ which are absent in the case of the real scalar. 

Let us carry out the program outlined in Sec.~\ref{sec:grav}. To integrate out the connection we derive the equations of motion for $v_\mu$, $a_\mu$, $\tau_{\m\n\rho}$ from the action~(\ref{eq:action_rho_f_1}) and solve them. Due to the presence of matter, the torsion components are expressed in terms of the derivative of $\Phi$ and the scalar and fermionic currents. Indeed, from 
\be
\f {\d S}{\d v_\m} = 0 \ ,~~~\f{\d S}{\d a_\m} = 0 \ ,~~~\f{\d S}{\d \tau_{\m\n\lambda}} = 0 \ ,
\ee
it is a straightforward computation to show that
\be
\label{eq:tors_eom}
M_P^2v_\m = \frac{-G_{aa}J^v_\m + G_{va}J^a_\m}{G_{vv}G_{aa}-G_{va}^2}\ ,~~~M_P^2a_\m = \frac{G_{va}J^v_\m - G_{vv}J^a_\m}{G_{vv}G_{aa}-G_{va}^2} \ ,~~~\tau_{\m\n\rho} = 0 \ ,
\ee
where we introduced the generalized ``currents''
\be
\label{eq:gener_curr}
J^{v/a}_\m = \partial_\mu Z^{v/a}_\Phi+Z^{v/a}_S S_\m + \z^{v/a}_V V_\m +\z^{v/a}_A A_\m \ .
\ee
Notice that the reduced torsion tensor $\tau_{\m\n\lambda}$ is zero on the equations of motion, an aftermath of the fact that, unlike $v_\m$ and $a_\m$, it is not sourced at this order in derivatives. Plugging Eq.~(\ref{eq:tors_eom}) into the action~(\ref{eq:action_rho_f_1}), we obtain\,\footnote{We omit the cosmological constant in what follows.}
\be
\begin{aligned}
\label{eq:action_gen_int-out_1}
S &=\int \diff^4 x \sqrt{g} \Bigg[ \f {M_P^2}{2}\Omega^2\mathring{R} - (\mathcal D_\m \Phi)^\ast \mathcal D^\m \Phi -U(\Phi^\ast\Phi)-\frac 1 4 F_{\m\n}^2 \\
&\quad\quad+\f i 2 \l( \overline{\Psi}\g^\m \mathring{D}_\m\Psi -  \overline{\mathring{D}_\m \Psi}\g^\m \Psi\r)-\frac{G_{aa}J^{v\,2}_\m+G_{vv}J^{a\,2}_\m-2G_{va}J^v_\m J^{a\,\m}}{2M_P^2\l(G_{vv}G_{aa}-G_{va}^2\r)}\Bigg] \ . 
\end{aligned}
\ee
At this point we use Eq.~(\ref{eq:gener_curr}) to express everything in terms of (the derivatives of) $\Phi$ and the currents $S,V,A$. Before we present the explicit result, it is convenient to move to the Einstein frame where the gravitational part of the action becomes canonical. We perform a Weyl rescaling of the metric
\be 
\label{eq:transform}
g_{\m\n} \mapsto \Omega^{-2} \tilde{g}_{\m\n} \ ,
\ee 
followed by a redefinition of the fermionic field
\be 
\label{eq:transform_fermions}
\Psi\mapsto \Omega^{3/2}\tilde{\Psi} \ , 
\ee
where we will omit the tilde in what follows. Then
we find
\be
\begin{aligned}
\label{eq:action_gen_int-out}
S &=\int \diff^4 x \sqrt{g} \Bigg[ \f {M_P^2}{2}\mathring{R} -  \f{1}{\Omega^2} (\mathcal D_\m \Phi)^\ast \mathcal D^\m \Phi - f(\Phi^\ast\Phi)- \frac{U(\Phi^\ast\Phi)}{\Omega^4}-\f 1 4 F_{\m\n}^2\\
&\qquad+\f i 2\l(  \overline{\Psi}\g^\m \mathring{D}_\m\Psi -   \overline{\mathring{D}_\m \Psi}\g^\m \Psi\r)+\frac{1}{M_P^2}\Big(\mathscr L _ {\Phi S}+\mathscr L _ {\Phi V} +\mathscr L _ {\Phi A}+\mathscr L _ {SS}\\
&\qquad\qquad\qquad\qquad\qquad\qquad\quad+\mathscr L _ {VV}+\mathscr L _ {AA}+\mathscr L _ {SV}+\mathscr L _ {SA}+\mathscr L _ {VA}\Big)\Bigg] \ .
\end{aligned}
\ee
Let us explain what are the different terms entering this expression. First, we recognize the standard Einstein-Hilbert term, the covariant kinetic and potential terms for $\Phi$ (rescaled by appropriate powers of the conformal factor), and the usual kinetic terms for the gauge and fermionic fields. Next, we have the function $f(\Phi^\ast\Phi)$ given by
\be
\label{eq:kh}
f(\Phi^\ast\Phi) =\Bigg( \frac{G_{aa}({Z^{v}_\Phi}')^{2}+G_{vv}({Z^{a}_\Phi}')^{2}-2G_{va}{Z^{v}_\Phi}'{Z^{a}_\Phi}'}{2M_P^2\Omega^2(G_{vv}G_{aa}-G_{va}^2)}+ \frac{3 M_P^2{\Omega'}^2}{\Omega^2} \Bigg)\p_\m(\Phi^\ast \Phi)\p^\m (\Phi^\ast \Phi) \;,
\ee
where prime stands for derivative with respect to $\Phi^*\Phi$. Being quadratic in the derivatives of the scalar, it contributes to its kinetic term; note, however, that it does not involve the covariant derivative. The rest of the terms describe various torsion-induced contact interactions between $\partial_\m(\Phi^*\Phi)$ and the scalar $S_\m$ and fermionic currents $V_\m, A_\m$. They read
\bea
\label{eq:LhS}
& &\mathscr L _{\Phi S} = \frac{1}{\Omega^2}\,\frac{G_{aa}Z^v_S {Z^{v}_\Phi}' +G_{vv}Z^a_S{Z^{a}_\Phi}'-G_{va}(Z^a_S{Z^{v}_\Phi}'+Z^v_S{Z^{a}_\Phi}')}{G_{va}^2-G_{vv}G_{aa}}\p_\m (\Phi^\ast\Phi) S^\m \ ,\\
& & \mathscr L _{\Phi V} = \frac{G_{aa}{Z^{v}_\Phi}'\z^v_V+G_{vv}{Z^{a}_\Phi}'\z^a_V-G_{va}(\z^v_V{Z^{a}_\Phi}'+{Z^{v}_\Phi}'\z^a_V)}{G_{va}^2-G_{vv}G_{aa}}\p_\m (\Phi^\ast\Phi) V^\m \ , \\
& & \mathscr L _{\Phi A} = \frac{G_{aa}{Z^{v}_\Phi}'\z^v_A+G_{vv}{Z^{a}_\Phi}'\z^a_A-G_{va}(\z^a_A{Z^{v}_\Phi}'+{Z^{a}_\Phi}'\z^v_A)}{G_{va}^2-G_{vv}G_{aa}}\p_\m (\Phi^\ast\Phi) A^\m \ ,\\
& & \mathscr L _{SS} = \frac{1}{\Omega^2}\,\frac{G_{aa}(Z_S^v)^2+G_{vv}(Z_S^a)^2-2G_{va}Z_S^v Z_S^a}{2\l(G_{va}^2-G_{vv}G_{aa}\r)}S_\m S^\m \ ,\\
& & \mathscr L _{VV} =\Omega^2\, \frac{G_{aa}(\z_V^v)^2+G_{vv}(\z_V^a)^2-2G_{va}\z_V^v \z_V^a}{2\l(G_{va}^2-G_{vv}G_{aa}\r)}V_\m V^\m \ , \label{eq:LVV}\\
& & \mathscr L _{AA} =\Omega^2\,\frac{G_{aa}(\z_A^v)^2+G_{vv}(\z_A^a)^2-2G_{va}\z_A^v \z_A^a}{2\l(G_{va}^2-G_{vv}G_{aa}\r)}A_\m A^\m \ , \label{eq:LAA}\\
& & \mathscr L _{SV} = \frac{G_{aa}Z^v_S\z^v_V+G_{vv}Z^a_S\z^a_V-G_{va}(Z^a_S\z^v_V+Z^v_S\z^a_V)}{G_{va}^2-G_{vv}G_{aa}}S_\m V^\m \ ,\\
& & \mathscr L _{SA} = \frac{G_{aa}Z^v_S\z^v_A+G_{vv}Z^a_S\z^a_A-G_{va}(Z^a_S\z^v_A+Z^v_S\z^a_A)}{G_{va}^2-G_{vv}G_{aa}} S_\m A^\m \ ,\\
\label{eq:LVA}
& & \mathscr L _{VA} = \Omega^2\,\frac{G_{aa}\z^v_V\z^v_A+G_{vv}\z^a_V\z^a_A-G_{va}(\z^a_V\z^v_A+\z^v_V\z^a_A)}{G_{va}^2-G_{vv} G_{aa}} V_\m A^\m \ .
\eea

Eqs.~(\ref{eq:action_gen_int-out})--(\ref{eq:LVA}) are the main results of the paper. Bearing in mind phenomenological applications, let us comment on how these considerations are applied to the SM. This is readily done by identifying the scalar field $\Phi$ with the Higgs doublet $H$, \ie replacing  $(\Phi,\Phi^\ast)\mapsto(H,H^\dagger)$, and requiring invariance under the electroweak group $SU(2)_{\rm L}\times U(1)_{\rm Y}$ instead of $U(1)$. Correspondingly, the scalar current becomes related to the hypercharge $U(1)_{\rm Y}$ and is given by
\be
S_\m =-\f i 2 \l( H^\dagger (\mathcal D_\m H)-(\mathcal D_\m H)^\dagger H\r) \ , 
\ee
where the covariant derivative is now $\mathcal D_\m = \p_\m -i g A_\m^{\rm a }T^{\rm a} -i\f{g'}{2} B_\m$ with ${\rm a}=1,2,3$ the $SU(2)_{\rm L}$ indices, while $g,g'$ and $A^{\rm a}_\m,B_\m$ are the couplings and gauge fields of the $SU(2)_{\rm L}$ and $U(1)_{\rm Y}$ groups, respectively. 
We notice that torsion induces interactions of the SM hypercharge current. Exploring their phenomenological consequences would be interesting and is left for future work. Note also that the $SU(2)_{\rm L}$ part of the electroweak group is not sensitive to torsion.

\subsection{Limiting cases}
\label{sec:lim}

If we do not impose criterion~\emph{iii)} from Sec.~\ref{sec:criteria}, the action~(\ref{eq:action_gen_int-out}) contains a functional freedom due to the various coefficient functions. In contrast, only a finite number of parameters are left once condition~\emph{iii)} is implemented and the functions are constrained according to Eq.~(\ref{eq:coefffunreal}) or (\ref{eq:coefffuncomplex}). Not counting the Planck mass, these are 3 in the gravity sector, 6 per real scalar field, 8 per complex scalar field and 4 per fermion.  In the following, we shall impose condition~\emph{iii)} and explore various relations between the action~(\ref{eq:action_gen_int-out}) and the models that have appeared previously in the literature. This will provide a useful check of our results. To the best of our knowledge,  the existing studies are limited to a real scalar field, so we replace $\Phi$ by $ \phi/\sqrt{2}$ and omit the gauge field and the current $S_\m$ in Eqs.~(\ref{eq:action_gen_int-out})--(\ref{eq:LVA}). 

\paragraph{Nonminimally coupled scalar field in Palatini gravity.}As explained in the introduction, the Palatini formulation of GR is the  limiting case of the EC theory in the absence of fermions. Models of Palatini gravity and a real scalar have been studied extensively \eg in the context of Higgs inflation~\cite{Bauer:2008zj}. We recover this theory by setting
\bea
c_{vv}= -\f 2 3 \ ,~~~c_{aa}=\f{1}{24} \ ,~~~\x_{vv} = \x_{aa}=-\z^v_\phi=\x\ ,
\eea
and equating the rest of the parameters to zero. For this choice, the terms in Eq.~(\ref{eq:kh}) cancel each other out, therefore
\be
f(\phi) = 0 \ .
\ee

\paragraph{Nonminimally coupled scalar field in metric gravity.}The metric formulation of the theory is restored for zero torsion. In this case the only nonvanishing term in Eq.~(\ref{eq:kh}) is the one coming from the Weyl transformation of the scalar curvature and is given by
\be
f(\phi)= \frac{3\x^2\phi^2}{M_P^2\Omega^4}(\partial \phi)^2 \ .
\ee
The is exactly the modification of the field's kinetic term in the original Higgs inflation model~\cite{Bezrukov:2007ep}. 

\paragraph{EC gravity with the Holst and Nieh-Yan terms.}In Ref.~\cite{Shaposhnikov:2020frq}, a generalization of the metric and Palatini scalar-gravity theories was suggested. It amounts to extending the EC action by coupling nonminimally a scalar field to the Einstein-Hilbert as well as the Holst and Nieh-Yan~\cite{Nieh:1981ww} invariants. The resulting theory was extensively studied in the context of inflation~\cite{Langvik:2020nrs, 2007.14978} and dark matter production~\cite{Shaposhnikov:2020aen}. The action~(\ref{eq:action_gen_int-out}) is an important step towards further generalizations of this model, which is reproduced by the following choice of parameters:
\be
\begin{aligned}
c_{vv} = -\frac{2}{3} \;,~ c_{va} = \frac{1}{6\bar\g}\;,~ & c_{aa}=\frac{1}{24}\;,~\x_{vv}=\x_{aa}=-\zeta^{v}_\phi=\x_h\;,~\x_{va}=\x_\gamma\;,~\zeta^{a}_\phi=\frac{\xi_\gamma}{4\bar{\gamma}}+\frac{\xi_\eta}{4} \;, \\
\zeta^v_V & =-\frac{\a}{2} \;,~\zeta^v_A=-\frac{\b}{2} \;, ~\zeta^a_V = 0\;,~\zeta^a_A=-\frac{1}{8}\;.
\end{aligned}
\ee
The torsion-induced dimension-six operators are in this case given by
\bea
& &f(\phi) =\f{3\phi^2}{\Omega^4 M_P^2}\f{ \l(\frac{\xi_\gamma -\xi_h}{\bar\g} + \xi_\eta \Omega^2\r)^2}{\Omega^4\l(\g^2(\phi)+1\r)} \partial_\mu\phi \partial^\mu\phi \ ,\\
& &\mathscr L_{h V} = \f{3\a}{4\Omega^2}\l(\x_h+\f{\g(\phi) }{\Omega^2(\g^2(\phi)+1)}\l(\frac{\xi_\gamma -\xi_h}{\bar\g} + \xi_\eta \Omega^2\r)\r)\p_\m \phi^2 V^\m\\
& &\mathscr L_{h A} = \f{3}{4\Omega^2}\l(\b\x_h+\f{1+\b\g(\phi) }{\Omega^2(\g^2(\phi)+1)}\l(\frac{\xi_\gamma -\xi_h}{\bar\g} + \xi_\eta \Omega^2\r)\r)\p_\m \phi^2 A^\m\\
& & \mathscr L_{VV} = \f{3\a^2}{16(\g^2(h)+1)}V_\m V^\m \ ,\\
& & \mathscr L_{AA} = \f{3\l(\b^2-2\b\g(\phi)-1\r)}{16(\g^2(\phi)+1)}A_\m A^\m \ ,\\
& & \mathscr L_{VA} = \f{3\a\l(\b-\g(h)\r)}{8(\g^2(\phi)+1)}V_\m A^\m \ ,
\eea
with 
\be
\g(\phi) = \f{1}{\bar\g \Omega^2}\l(1+\f{\x_\g \phi^2}{M_P^2}\r) \ .
\ee

\section{A glimpse on curvature-squared terms}
\label{sec:curvatureSquare}

The action~(\ref{eq:action_gen_int-out}) contains all terms compatible with the two criteria from Sec.~\ref{sec:criteria}. Recall that the rationale behind~\emph{i)} was to ensure that the particle spectrum is the same as in GR in the metric formulation. In particular, condition~\emph{i)} excludes all higher-curvature invariants. Although such terms usually bring new degrees of freedom,  this is not always the case. In order to illustrate this, let us construct a specific example of a model with a curvature-squared term that propagates only the massless graviton. The starting point is
\be \label{eq:Sprime}
S' = \int \diff^4 x\sqrt g\Bigg[M_P^2 F +c F^2 +j^v_\m v^\m +j^a_\m a^\m \Bigg] +S_m\ ,
\ee
where $c$ is a constant, and the curvature $F$ is defined in \Eq~(\ref{eq:f_def}). We allowed for arbitrary couplings of the torsion vector and pseudovector to the currents $j^{v/a}_\m$ that depend on matter fields, thus without loss of generality we take the matter action $S_m$ not to contain torsion.  

The reason we did not include torsion-squared terms in $S'$, apart from those already contained in $F$, is twofold. First, introducing $a_\m^2$ or $\tau_{\kappa\lambda\m}^2$ does not affect our conclusions, so we omit them for simplicity. Second, introducing $v_\m^2$ and/or $v_\m a^\m$ would actually modify the dynamics of the theory so that it propagates a new scalar degree of freedom in addition to the graviton. We will comment on this in what follows.

To proceed, we rewrite Eq.~(\ref{eq:Sprime}) in a more convenient form by introducing a Lagrange multiplier $\lambda=\lambda(x)$, so that
\be
S'~~\mapsto~~S = \int \diff^4 x\sqrt g\Bigg[ (M_P^2+4c\lambda) F -4c\lambda^2 +j^v_\m v^\m +j^a_\m a^\m \Bigg]+S_m \ . 
\ee
It is clear that $S=S'$ on $\lambda$'s equation of motion. Decomposing $F$ as in Eq.~(\ref{eq:f_decomp}), solving for torsion and performing a Weyl rescaling of the metric with conformal factor\,\footnote{The order of operations is not essential.} 
\be
\Omega^2 = 1+ \f{4c\lambda}{M_P^2} \ ,
\ee
we arrive at
\be \label{eq:MetricTheoryCurvatureSquare}
\widetilde{S} = \int \diff^4x \sqrt{\tilde{g}} \Bigg[\f{M_P^2}{2}\widetilde{\mathring R} -\f{M_P^4}{4c}\l(1-\Omega^{-2}\r)^2 +\f{3\l(\tilde{j}^v_\m\r)^2}{4M_P^2\,\Omega^4}-\f{12\l(\tilde{j}^a_\m\r)^2}{M_P^2\,\Omega^4}+\f{3}{2}\p_\m \Omega^{-2}\tilde{j}^{v\,\m}\Bigg] +\widetilde{S}_m\ , 
\ee
where the tilde denotes the Weyl-transformed quantity. In general, the matter currents $\tilde{j}^{v/a}_\mu$ as well as the action $\widetilde{S}_m$ can have an explicit dependence on $\Omega$ and, hence, on $\lambda$. Nevertheless, irrespective of how the Lagrange multiplier enters the action, no kinetic term for it can be generated provided that we start from $S'$ or $S$.\footnote{This is ultimately due to the fact that $F$ transforms covariantly under Weyl rescalings.} This would not have been the case had we included $v_\m v^\m$ and/or $v_\m a^\m$ in the action, since the equation of motion for torsion would have acquired pieces $\propto \p_\m \Omega$, rendering the field dynamical.

If we momentarily neglect $\widetilde{S}_m$, we can integrate out $\lambda$ or, equivalently, $\Omega$, via its equation of motion. This gives
\be 
\label{eq:omega_higher_curv}
 \Omega^2 = \frac{M_P^6 + 3 c \l( 16 (\tilde{j}^a_{\mu})^2 - (\tilde{j}^v_{\mu})^2 \r) }{M_P^6 - 3c M_P^2 \tilde{\nabla}_\mu \tilde{j}^{v\,\m} } \;.
\ee
First, we observe that in the absence of external currents, $\tilde{j}^v_\m = \tilde{j}^a_\m=0$, the result is $\Omega^2=1$ and the action \eqref{eq:MetricTheoryCurvatureSquare} reduces to metric gravity. On the other hand, if the currents are non-vanishing, then plugging $\Omega^2$ in \Eq~\eqref{eq:MetricTheoryCurvatureSquare} leads to a series of nontrivial higher-order operators. In general, they are different from the ones in \Eq~\eqref{eq:action_gen_int-out} that are obtained from the linear-in-curvature terms only. This is already clear from the presence of covariant divergence of $\tilde j^v_\m$ in Eq.~(\ref{eq:omega_higher_curv}). Of course, if we expand this equation in powers of $M_P^{-2}$, the form of the leading,  dimension-six, operators will be the same as in \Eq~(\ref{eq:action_gen_int-out}).

To summarize, it is possible to come up with ``fine-tuned'' models with curvature-squared terms that do not propagate new gravitational degrees of freedom and lead to nontrivial contact interactions in their metrical form. However, such theories are by no means generic. Nevertheless, it would be interesting to study them systematically. 

What if one allows for extra degrees of freedom of gravitational origin? For concreteness, let us focus again on the curvature-squared operators. As discussed above, such operators would in general bring about ghosts and/or tachyons. This is partially due to higher-derivatives, but also in the absence of those the Poincar\'e group, being noncompact, can in general not ensure positive-definiteness of all kinetic and mass terms. Again, there are exceptions to this expectation---higher-derivative theories are not necessarily plagued by inconsistencies, see~\cite{Stelle:1977ry,Neville:1978bk,Hayashi:1979wj,*Hayashi:1980qp,Sezgin:1979zf,Sezgin:1981xs,Kuhfuss:1986rb,Yo:1999ex,Yo:2001sy,Nair:2008yh,Nikiforova:2009qr,Karananas:2014pxa,*Karananas:2016ltn,Obukhov:2017pxa,Blagojevic:2017ssv,Blagojevic:2018dpz,Lin:2018awc,Jimenez:2019qjc,Lin:2019ugq,Percacci:2019hxn} for a non-exhaustive list of references,  as well as~\cite{Lin:2018awc} and~\cite{Blagojevic:2018dpz} for the most recent and complete analyses of the quadratic parity-preserving and parity-violating Poincar\'e gauge theory, respectively. Perhaps, the most well-known and studied example of a healthy theory is given by the following action (see \eg~\cite{Stelle:1977ry,Starobinsky:1980te})
\be \label{eq:R2}
\int\diff^4x\sqrt{g}\l(\f{M_P^2}{2}\mathring R +c\mathring R^2\r) \ ,
\ee
that propagates a massive spin-0 particle in addition to the graviton.\footnote{One can introduce an additional nonminimally coupled scalar playing the role of the Higgs field. The resulting model was proposed  in~\cite{Gorbunov:2018llf} in the context of Higgs inflation.} Speaking more generally, it is possible to eliminate higher derivatives by having the curvature-squared terms combine in very specific ways. To give an idea, two such examples are the following
\bea
\label{eq:higher_curv_1}
{\rm C}_1&=&F_{ABCD}F^{ABCD}-2F_{ABCD}F^{CDAB}+2 F_{ABCD}F^{ACBD}  \ ,
\\
\label{eq:higher_curv_2}
{\rm C}_2&=&F_{ABCD}F^{ABCD}+F_{ABCD}F^{CDAB}-4 F_{ABCD}F^{ACBD}  \ ,
\eea
with $F_{ABCD}=e^\m_A e^\n_B \eta_{CI}\eta_{DJ}F^{IJ}_{\m\n}$. It can be readily checked that, upon decomposing the connection as in Eq.~(\ref{eq:conn_split}), both expressions read, schematically, 
\be
{\rm C}_1,{\rm C}_2\supset{\rm Riemann} ^ 2 +  {\rm Riemann}\times \partial({\rm Torsion})+\ldots  \ ,
\ee
where the ellipses stand for terms with at most two derivatives. Working at the level of the action, after some integrations by parts and using the algebraic and differential Bianchi identities obeyed by $\mathring R_{\kappa\lambda\m\n}$, both the ${\rm Riemann}^2$- and the ${\rm Riemann}\times \partial({\rm Torsion})$-contributions are found to vanish in~(\ref{eq:higher_curv_1}) and~(\ref{eq:higher_curv_2}).

\section{Discussions and Outlook}
\label{sec:conclusion}

The Einstein-Cartan formulation of General Relativity naturally arises in the gauge approach to gravity. This motivates a close study of this theory from both theoretical and phenomenological perspectives. In particular, it is important to understand how it deviates from the metric formulation of GR, which is most commonly used.
The important new ingredient in this incarnation of gravity is spacetime torsion. In the absence of matter, torsion is not sourced, resulting into the EC and metric formulations being (at least classically) completely equivalent.\footnote{Palatini gravity, which represents a special case of EC theory, has been proven to be equivalent to the metric formulation also on the quantum level \cite{Aros:2003bi}.} This changes once matter is introduced, and in this paper we painted a quantitative picture of the differences. 

To this end, we first devised criteria for coupling the SM to gravity in a generic way. We required that the admissible terms are at most quadratic in the derivatives and of mass dimension not bigger than four, so that we were restricted to terms at most linear in curvature and quadratic in torsion.\footnote{Already at mass dimension five, many more terms appear. Some of them, as well as their implications for low-energy phenomenology, were \eg discussed in \cite{Kostelecky:2007kx, Obukhov:2014fta}.} This was sufficient to exclude additional propagating degrees of freedom beyond the massless graviton and those already present in the matter sector. Subsequently, we constructed an action for EC gravity coupled to matter, where we took into account fermions, real or complex scalars and gauge bosons, and included all terms that fulfill the criteria devised before.

In our setup, the connection and consequently torsion are manifestly nondynamical. Nevertheless, its presence affects nontrivially the dynamics of the rest of the fields. To explicitly illustrate this, we constructed an equivalent torsion-free theory by eliminating the connection via its equation of motion. Various higher-dimensional terms describing interactions between matter currents and field derivatives appear this way, see Eqs.~(\ref{eq:kh})-(\ref{eq:LVA}). They comprise four-fermi interactions, mixing between scalar and fermionic currents and a modified kinetic term for the scalar(s). In general, the interactions come with arbitrary field-dependent coefficient functions. We can reduce the functional freedom contained in them to a finite number of parameters by restricting ourselves to nonminimal couplings of the form displayed in \Eqs~(\ref{eq:coefffunreal}) and~(\ref{eq:coefffuncomplex}). However, our analysis is valid beyond such a restriction. In any case, the torsion-induced operators form a subset of the plethora of possible higher-dimensional terms that we would have added in an effective field theory approach that starts directly from the metric theory.

An interesting novel result of our analysis concerns gauge theories. Given a scalar current associated with an Abelian (sub)group, it can and actually should be coupled to torsion in all possible manners. Consequently, the effective metric theory contains interactions of this current with itself and the rest of the fields. We demonstrated how this happens in an Abelian toy model. If the matter sector is identified with the Standard Model, it is the radial mode of the Higgs field that we envisage being nonminimally coupled to the various geometrical objects. The aforementioned scalar current then corresponds to the hypercharge and may give rise to interesting physics.

It is important to note that the theory we constructed has many free parameters and coefficient functions. This means that the mass scales suppressing the higher-dimensional operators are, in principle, field-dependent and arbitrary. A judicial choice of parameters can make the suppression scale lower than the Planck mass and lead to a rich phenomenology. For example, it provides a mechanism for producing singlet fermions, which can assume the role of dark matter, in the early Universe~\cite{Shaposhnikov:2020aen}.

As discussed above, confining ourselves to terms linear in curvature is a sufficient but not a necessary condition for the absence of extra gravitational degrees of freedom. Thus, in principle, this requirement can be relaxed by including curvature-squared terms in the action. At the same time, this is something that has to be done with extra care, since quite generically, higher-curvature invariants are intrinsically linked to  pathologies. In other words, only particular combinations of curvature-squared invariants should be allowed in the action for the theory to be healthy. We presented a corresponding example in Sec.~\ref{sec:curvatureSquare}. It would be interesting to systematically study the implications of such modifications for the higher-dimensional interactions between the matter fields.

Let us conclude by discussing two distinct but interrelated points. The first is how to reduce the arbitrariness of the theory that we constructed here. An appealing way to achieve that is by Weyl-gauging the action, \ie requiring it to be invariant under gauged dilatations. We will elaborate on this in~\cite{scale_Weyl_EC}, but we shall give a brief outlook already here.
Although invoking the gauge principle to constrain a theory without altering its spectrum may sound counterintuitive at first sight, in the sense that it normally necessitates the introduction of extra dynamical degrees of freedom, this is not the case for spacetime symmetries. The reason is that certain geometrical quantities present in the theory we constructed, more specifically the Ricci scalar and torsion vector, transform inhomogeneously under Weyl rescalings. Both can therefore assume the role of effective gauge fields and compensate for the inhomogeneous pieces coming from their own transformations as well as the ones from the kinetic term of the scalar. This means that in order for the action to exhibit Weyl invariance, the coefficient functions of these terms are not free anymore, but are rather related to each other. Actually, this is a direct generalization of what happens with the conformally coupled scalar field in conventional GR; there, the nonminimal coupling of the field is fixed by conformal symmetry to be equal to the well known value $-1/6$ (in our conventions). The similarities do not end here: the Weyl redundancy of the action is actually too much, since it translates into $\phi$ being spurious also in the EC case. Thus, to maintain a propagating spin-0 degree of freedom in the spectrum, it is unavoidable to introduce yet another scalar field and have the gauge freedom eliminate this instead.  

What was implicitly assumed in the above discussion is that the starting point for Weyl-gauging is a biscalar theory invariant under global dilatations. This brings us to the second point. The metric counterpart of exactly this, globally scale-invariant, theory is  the ``Higgs-dilaton model'' introduced in~\cite{Shaposhnikov:2008xb}. It is an economic and, at the same time, phenomenologically viable scale-invariant extension of the Standard Model plus GR that has been extensively studied and generalized, see~\cite{Shaposhnikov:2008xi,Shaposhnikov:2008ar,Shaposhnikov:2009nk,Blas:2011ac,GarciaBellido:2011de,GarciaBellido:2012zu, Bezrukov:2012hx,Armillis:2013wya,Gretsch:2013ooa,Rubio:2014wta,Ghilencea:2015mza,Ferreira:2016vsc,Karananas:2016kyt,Ferreira:2016wem,Ferreira:2016kxi,Ghilencea:2016dsl,Shaposhnikov:2018jag,Casas:2018fum,Shkerin:2019mmu,Herrero-Valea:2019hde,Rubio:2020zht}~for a far-from-complete list of references. In the cosmological context, the model predicts a rather interesting phenomenology for the early and late Universe, in complete agreement with observations. Moreover, it connects these eras via a set of consistency conditions between inflationary and dark energy observables~\cite{GarciaBellido:2011de,Casas:2017wjh}, which will hopefully be testable in the near future. It would be interesting to understand which of the attractive features of the metrical Higgs dilaton model survive when generalized this way. As far as particle physics is concerned, scale (and conformal) symmetry may be relevant for addressing the fine-tuning issues of the Standard Model~\cite{Wetterich:1983bi,Bardeen:1995kv,Shaposhnikov:2007nj,Shaposhnikov:2018xkv,Mooij:2018hew,Shaposhnikov:2018nnm,Wetterich:2019qzx,Shaposhnikov:2020geh,Karananas:2020qkp}. Given the prominent role that gravity plays in both the hierarchy and cosmological constant puzzles, it is certainly worth studying if and what changes when the gravitational dynamics is described in terms of the  Einstein-Cartan theory.

\section*{Acknowledgments} 

This work was supported by the ERC-AdG-2015 grant 694896 and by the Swiss National Science Foundation Excellence grant 200020B\_182864. The work of A.~S. was in part supported by the Department of Energy Grant DE-SC0011842.

{\small
\setlength{\bibsep}{1.3pt plus 0.1ex}
\bibliographystyle{utphys}
\bibliography{Nonpropagating_Torsion_Higgs.bib}
}

\end{document}